\documentclass[12pt]{article}
\usepackage{hyperref,amsfonts,amsmath,amssymb,graphicx,bm,subfigure,a4wide}
\begin{document}

\title{\textbf{Influence of the hypermagnetic field noise on the baryon asymmetry generation in the symmetric phase of the early universe}}

\author{Maxim Dvornikov\thanks{maxdvo@izmiran.ru},\ 
Victor B. Semikoz\thanks{semikoz@yandex.ru}
\\
\small{\ Pushkov Institute of Terrestrial Magnetism, Ionosphere} \\
\small{and Radiowave Propagation (IZMIRAN),} \\
\small{108840 Moscow, Troitsk, Russia}}

\date{}

\maketitle

\begin{abstract}
We study a matter turbulence caused by strong random hypermagnetic fields (HMFs ) that influence the baryon asymmetry evolution  due to the Abelian anomalies in the symmetric phase in the early Universe. Such a matter turbulence is stipulated by the presence of the advection term in the induction equation for which a fluid velocity is dominated by the Lorentz force in the Navier-Stokes equation. For random HMFs, having nonzero mean squared strengths, we calculate the spectra for the HMF energy and the HMF helicity densities. The latter function governs the evolution of the fermion asymmetries in the symmetric phase before the electroweak phase transition (EWPT). In the simplest model based on the first SM generation for the lepton asymmetries of $e_\mathrm{R,L}$ and $\nu_{e_\mathrm{L}}$, we calculate a decline of all fermion asymmetries including the baryon asymmetry, given by the `t~Hooft conservation law, when one accounts for a turbulence of HMFs during the universe cooling down to EWPT. We obtain that the stronger the mean squared strength of random initial HMFs is, the deeper the fermion asymmetries decrease, compared to the case in the absence of any turbulence.
\end{abstract}

\section{Introduction\label{sec:INTR}}

The origin of the baryon asymmetry of the Universe (BAU) is a long standing issue. One of the scenarios for the baryogenesis implies the production of a lepton asymmetry first. Then, this lepton asymmetry is converted into BAU owing to the $B/3-L$ conservation in the standard model (SM). The numerous models of the leptogenesis are reviewed in Ref.~\cite{BucPecYan05}.

Many models of the leptogenesis are based on the particle physics beyond SM, where the lepton number is violated generically. There is, however, another approach which implies the presence of helical hypermagnetic fields (HMFs) in plasma of the early Universe
before the electroweak phase transition (EWPT). The nonzero hypermagnetic helicity affects the lepton asymmetry evolution owing to the analog of the Adler anomaly for HMFs. The lepton asymmetry, in its turn, contributes the hypermagnetic fields evolution owing to the analog of the chiral magnetic effect (the CME)~\cite{Vilenkin:1980fu}. One has to account for chirality flip processes due to direct (inverse) Higgs decays and sphaleron transitions which violate the left lepton number and wash out  BAU.

The scenario of the baryogenesis, previously proposed in Ref.~\cite{Davidson}, assumes the presence of a nonzero right electron asymmetry ${\rm L}_{e\mathrm{R}}(T)=(n_{e\mathrm{R}} - n_{\bar{e}\mathrm{R}})/s\neq 0$ long before the electroweak phase transition (EWPT), $T> T_\mathrm{RL}\simeq 10\,\mathrm{TeV}\gg T_\mathrm{EWPT}\simeq 100\,\mathrm{GeV}$. Here $s$  is the entropy density and $n_{(e,\bar{e})\mathrm{R}}$ are the number densities of right electrons and positrons. In this case, the influence of sphalerons, that could wash out BAU, ${\rm B}=(n_{\rm B} - n_{\bar{{\rm B}}})/s\to 0$, was minimized since sphalerons are left chirality objects. 

As we mentioned above, the observed value of  ${\rm BAU}\simeq 10^{-10}$, can be produced in HMF before EWPT, $T\geq  T_\mathrm{EWPT}$. This possibility was first proposed in Refs.~\cite{Shaposh1,Shaposh2}. These fields are the drivers of the leptogenesis through Abelian anomalies for HMF. The growth of the total lepton number, e.g., for the first generation ${\rm L}_e={\rm L}_{e\mathrm{R}} + {\rm L}_{e\mathrm{L}} + {\rm L}_{\nu_{e\mathrm{L}}}$, where ${\rm L}_a=(n_a -n_{\bar{a}})/s$, means the growth of the BAU due to the `t~Hooft's conservation law for the baryon number ${\rm B}- 3{\rm L}_e={\rm const}$. 

In the scenario, where the right electron asymmetry supports the
BAU alone at temperatures $T > T_\mathrm{RL}\simeq 10\,{\rm TeV}$, the
following universe cooling leads to a nonzero mixing between the left and right lepton 
asymmetries of electrons and neutrinos. It happens since Higgs bosons decays become faster than the universe expansion, $\Gamma_\mathrm{RL}\sim T > \mathrm{H}\sim T^2$.
Thus, the production of left leptons starts at $T = T_\mathrm{RL}$.
It results in the self-consistent evolution of the
right and the left electron asymmetries at $T < T_\mathrm{RL}$ through the corresponding Abelian anomalies in
SM in the presence of a seed HMF~\cite{DS3}. We choose below $t_0=(2{\rm H})^{-1}=\tilde{M}_\mathrm{Pl}/2T_\mathrm{RL}^2$  as the initial time in our problem, where $\tilde{M}_\mathrm{Pl}=7\times 10^{17}\,\mathrm{GeV}$ is the effective Planck mass and $\mathrm{H}$ is the Hubble parameter. 

The hypermagnetic helicity evolution proceeds in
a self-consistent way with the lepton asymmetry growth. The role of sphaleron transitions, decreasing
the left lepton number, turns out to be negligible in given scenario. The hypermagnetic helicity plays
a key role in lepto- and baryogenesis in our scenario. The closer HMF to a maximally helical one is, the faster BAU grows up the observable value, $(\text{BAU})_\text{obs}\simeq 10^{-10}$.

Although the precise estimate of BAU can be made only by solving all evolution
equations for the all fermion and the Higgs boson asymmetries numerically (see, e.g., Refs.~\cite{FK,Kamada}), 
the qualitative behavior of the BAU evolution in HMFs can
be more easily understood in the simplified model developed in Ref.~\cite{DS3}. For example, in Ref.~\cite{DS3}, we could account for the correction to the helicity parameter $\alpha_\mathrm{Y}$, associated with the left fermions asymmetry (see Eq.~(\ref{Faraday-parameters}) below). Recently, the BAU production was discussed in Ref.~\cite{Abbaslu}, where the temperature range $100\,\mathrm{GeV}\leq T\leq 10\,\mathrm{TeV}$ was considered. Moreover, the configuration of HMF which includes both helical ($B_a, B_b$) and non-helical ($B_z$) components, as well as the plasma vorticity, were accounted for in Ref.~\cite{Abbaslu} to follow the evolution of the matter-antimatter asymmetry in the symmetric phase of the early Universe.

The goal of the present work is the study of the influence of the matter noise, or the turbulent motion of plasma, to the evolution of BAU 
and HMFs, both the HMF energy and the HMF helicity densities, in the symmetric phase before EWPT. The turbulent matter motion and random HMFs, having $\langle {\bf B}_\mathrm{Y}\rangle=0$, are related to each other. We rely below on the simplified solution of the Navier-Stokes equation for the fluid velocity ${\bf v}$ as suggested, e.g., in Ref.~\cite{Sigl:2002kt},
\begin{equation}\label{drag}
  {\bf v}= \frac{\tau_d(\nabla\times {\bf B}_\mathrm{Y})\times{\bf B}_\mathrm{Y}}{p + \rho}.
\end{equation}
It allows one to express the advection (dynamo) term $\nabla\times ({\bf v}\times {\bf B}_\mathrm{Y})$ in the induction (Faraday) equation using the explicit form in Eq.~(\ref{drag}) through HMFs. Here $p=\rho/3$ is the equation of state in relativistic plasma, $\tau_d\simeq l_\mathrm{free}=(\sigma_\mathrm{coll}n)^{-1}$ is the drag time  given by the free path for leptons (including neutrinos) that interact each other via ``Coulomb" collisions, $\sigma_\mathrm{coll}\simeq \alpha'^{2}/T^2$ is the corresponding cross-section~\cite{fn1}, $\alpha'=g'^{2}/4\pi=(137\cos^2 \theta_\mathrm{W})^{-1}$ is the analog of the fine structure constant given by the hypercharge $g'=e/\cos \theta_\mathrm{W}$ in SM, and $n\sim T^3$ is the particle density.

In a strong HMF, the Lorentz force ${\bf F}_\mathrm{L}\simeq (\nabla\times {\bf B}_\mathrm{Y})\times {\bf B}_\mathrm{Y}$ prevails other terms in the Navier-Stokes equation. The Larmor period becomes much shorter than the drag time, $T_\mathrm{Larm}\ll \tau_d$. It means that the matter fluid is accelerated $\partial_t{\bf v}\sim {\bf F}_\mathrm{L}$ before any Coulomb collision happens.

Our paper is organized as follows. In Sec.~\ref{sec:MATTNOISE}, we consider the matter noise influence via Eq.~(\ref{drag}) for the spectra of HMF energy and helicity densities in the symmetric phase of the early Universe. In Sec.~\ref{sec:KINEQS}, we reconsider the kinetic equations for the particle density asymmetries based on the first lepton generation in SM that are valid in the symmetric phase at $T> T_\mathrm{EWPT}$. In Sec.~\ref{sec:MU5}, we check the validity of a marked CME imbalance $\mu_5(\eta)=[\xi_{e\mathrm{R}}(\eta) - \xi_{e\mathrm{L}}(\eta)]/2\neq 0$ coming from the symmetric phase just before the EWPT. In Sec.~\ref{sec:COMPLKINEQS}, we complete the set  of the kinetic equations for the HMF spectra, the lepton and Higgs boson asymmetries. In Sec.~\ref{sec:INICOND}, we formulate the initial condition for the derived kinetic equations. In Sec.~\ref{sec:BAUXIER}, we explore the noise matter influence both the BAU and lepton asymmetries. Finally, we discuss the validity of our approach for the BAU generation within SM in Sec.~\ref{sec:DISC}. 

\section{Matter noise influence the evolution for spectra of HMF energy and helicity densities\label{sec:MATTNOISE}}

We start from the induction (Faraday) equation for the HMF ${\bf B}_\mathrm{Y}$,
\begin{equation}\label{Faraday}
  \frac{\partial {\bf B}_\mathrm{Y}}{\partial t} =
  \nabla\times {\bf v}\times {\bf B}_\mathrm{Y} + \gamma_\mathrm{D}\nabla^2{\bf B}_\mathrm{Y} +
  \gamma_{\omega}\nabla\times \bm{\omega} + \alpha_\mathrm{Y}\nabla\times {\bf B}_\mathrm{Y},
\end{equation}
where $\bm{\omega}=\nabla\times {\bf v}$ is the fluid vorticity, the coefficients
\begin{equation}\label{Faraday-parameters}
  \gamma_\mathrm{D}=\frac{1}{\sigma_\mathrm{cond}},
  \quad
  \gamma_{\omega}=\frac{g'}{8\pi^2\sigma_\mathrm{cond}}(\mu_{e\mathrm{R}}^2 - \mu_{e\mathrm{L}}^2),
  \quad
  \alpha_\mathrm{Y}=\frac{g'^{2}}{4\pi^2\sigma_\mathrm{cond}}
  \left(
    \mu_{e\mathrm{R}} - \frac{\mu_{e\mathrm{L}}}{2}
  \right),
\end{equation}
are the functions of time during the radiation era, $t=(2{\rm H})^{-1}=\tilde{M}_\mathrm{Pl}/2T^2$, $\sigma_\mathrm{cond}=\sigma_cT\simeq 100 T$ is the plasma conductivity. The analogue of Eq.~\eqref{Faraday} for Maxwell's fields is given in Ref.~\cite{Tashiro:2012mf}.

In order to derive evolution equations for the binary functions: (i) the HMF energy  density $\rho_{\mathrm{B}_\mathrm{Y}}\sim B_\mathrm{Y}^2$ and (ii) the HMF helicity density $h_\mathrm{Y}\sim {\bf Y}\cdot{\bf B}_\mathrm{Y}$ where 
${\bf B}_\mathrm{Y}=\nabla\times {\bf Y}$, we use also the corresponding equation for the potential ${\bf Y}$,
\begin{equation}\label{potential}
  \frac{\partial {\bf Y}}{\partial t} = {\bf v}\times {\bf B}_\mathrm{Y} - \gamma_\mathrm{D}\nabla\times {\bf B}_\mathrm{Y} +
  \gamma_{\omega}\bm{\omega} + \alpha_\mathrm{Y}{\bf B}_\mathrm{Y}.
\end{equation}
Using Eqs.~(\ref{Faraday}) and~(\ref{potential}), one can obtain the
evolution equations for the real binary products in
the Fourier representation, $\partial_t\mathcal{E}_{\rm B_\mathrm{Y}}(k, t)\sim [\dot{\bf B}_\mathrm{Y}(k,t)\cdot{\bf B}^*_\mathrm{Y}(k,t) + \dot{\bf B}^*_\mathrm{Y}(k,t)\cdot{\bf B}_\mathrm{Y}(k, t)]$ where
\begin{equation}
  \mathcal{E}_{\mathrm{B}_\mathrm{Y}}(t) =
  \frac{1}{2V}
  \int \frac{\mathrm{d}^3k}{(2\pi)^3}
  |{\bf B}_\mathrm{Y}(k,t)|^2= \int \mathrm{d}k\mathcal{E}_{B_\mathrm{Y}}(k,t) = B_\mathrm{Y}^2(t)/2
\end{equation}
is the hypermagnetic energy density, and
$\partial_t \mathcal{H}_\mathrm{Y}(k,t)\sim [\dot{{\bf Y}}(k,t)\cdot{{\bf B}}_\mathrm{Y}^*(k,t) + {\bf Y}(k,t)\cdot \dot{{\bf B}}_\mathrm{Y}^*(k,t)]$, where
\begin{equation}
  \mathcal{H}_\mathrm{Y}(t) = \frac{1}{V}
  \int \frac{\mathrm{d}^3k}{(2\pi)^3}
  [{\bf Y}(k,t)\cdot{\bf B}_\mathrm{Y}^*(k,t)] =
  \int \mathcal{H}_\mathrm{Y} (k,t)\mathrm{d}k
\end{equation}
is the hypermagnetic helicity density. We
use below the conformal variables with the time $\eta= \tilde{M}_\mathrm{Pl}/T$,
where $\tilde{M}_\mathrm{Pl} = M_\mathrm{Pl}/1.66\sqrt{g_*}$, $M_\mathrm{Pl} = 1.2\times 10^{19}\,\mathrm{GeV}$ is
the Planck mass, $g_* = 106.75$ is the number of relativistic
degrees of freedom in the hot plasma before EWPT.

In order to get the kinetics for the conformal HMF spectra, $\mathcal{\tilde{H}}_{\rm B_\mathrm{Y}}\equiv\mathcal{\tilde{H}}_{\rm B_\mathrm{Y}}(\tilde{k},\eta)$ and $\mathcal{\tilde{E}}_{\rm B_\mathrm{Y}}\equiv\mathcal{\tilde{E}}_{\rm B_\mathrm{Y}}(\tilde{k},\eta)$,
we modify the system in Eq.~(3.11) in Ref.~\cite{DS} written there for Maxwell's fields in the broken phase, $T<T_\mathrm{EWPT}$, to the case of hypermagnetic fields in the symmetric phase, $T>T_\mathrm{EWPT}$,
\begin{align}\label{spectra}
  \frac{\partial \mathcal{\tilde{E}}_{\rm B_\mathrm{Y}}}{\partial \eta} = &
  - 2\tilde{k}^2\tilde{\eta}_\mathrm{eff}\mathcal{\tilde{E}}_{\rm B_\mathrm{Y}} +
  \tilde{\alpha}_+\tilde{k}^2\mathcal{\tilde{H}}_{\rm B_\mathrm{Y}},
  \nonumber
  \\
  \frac{\partial \mathcal{\tilde{H}}_{\rm B_\mathrm{Y}}}{\partial \eta}= &
  - 2\tilde{k}^2\tilde{\eta}_\mathrm{eff}\mathcal{\tilde{H}}_{\rm B_\mathrm{Y}} +
  4\tilde{\alpha}_-\mathcal{\tilde{E}}_{\rm B_\mathrm{Y}},
\end{align}
where  $\mathcal{\tilde{E}}_{\rm B_\mathrm{Y}}=a^3\mathcal{E}_{\rm B_\mathrm{Y}}$ and $\mathcal{\tilde{H}}_{\rm B_\mathrm{Y}}=a^2\mathcal{H}_{\rm B_\mathrm{Y}}$ are the dimensionless spectra.

The dimensionless kinetic coefficients in Eq.~\eqref{spectra}, $\tilde{\eta}_\mathrm{eff}(\eta)$ is the diffusion one and $\alpha_\mathrm{Y}(t)=\Pi(t)/\sigma_\mathrm{cond}$ is the hypermagnetic helicity coefficient, are modified due to the presence of a matter noise,
\begin{align}\label{parameters}
  \tilde{\eta}_\mathrm{eff}=&
  \frac{\eta_\mathrm{eff}}{a}=\sigma_c^{-1} +
  \frac{4}{3}\frac{(\alpha')^{-2}}{\tilde{\rho} + \tilde{p}}
  \int_{\tilde{k}_\mathrm{min}}^{\tilde{k}_\mathrm{max}} \mathrm{d}\tilde{k}\mathcal{\tilde{E}}_{{\rm B_\mathrm{Y}}},
  \quad
  \alpha'=\frac{g'^{2}}{4\pi},
  \nonumber
  \\
  \tilde{\alpha}_{\pm}=&\alpha_{\pm}=\alpha_\mathrm{Y}(\eta) \mp \frac{2}{3}\frac{(\alpha')^{-2}}{\tilde{\rho} +
  \tilde{p}}\int_{\tilde{k}_\mathrm{min}}^{\tilde{k}_\mathrm{max}} \mathrm{d}\tilde{k}\tilde{k}^2\mathcal{\tilde{H}}_{{\rm B_\mathrm{Y}}}.
\end{align}
Here we changed the pseudoscalar helicity parameter in CME for Maxwellian fields applied in Ref.~\cite{DS},
$\alpha_{\rm CME}(t)=\Pi(t)/\sigma_\mathrm{cond}=2\alpha_\mathrm{em}\mu_5(t)/\pi\sigma_\mathrm{cond}$, to the scalar helicity parameter $\alpha_\mathrm{Y}$ for HMFs in the symmetric phase,\footnote{In fact, we change the total helicity parameter $\alpha_\mathrm{Y}(\eta)=\alpha'/(\pi\sigma_c)\left(\xi_{e\mathrm{R}}(\eta) - \xi_{e\mathrm{L}}(\eta)/2 + 3\xi_\mathrm{B}(\eta)/2\right)$ derived in Ref.~\cite{Kamada} neglecting a small baryon contribution compared to the lepton asymmetries, $\xi_\mathrm{B}\ll \xi_{e\mathrm{R},e\mathrm{L}}$}
\begin{equation}\label{alpha}
  \alpha_\mathrm{Y}(\eta)=\frac{\Pi(t)}{\sigma_\mathrm{cond}}=
  \frac{\alpha'}{\pi\sigma_c}
  \left[
    \xi_{e\mathrm{R}}(\eta) - \frac{\xi_{e\mathrm{L}}(\eta)}{2}
  \right].
\end{equation}
In Eq.~(\ref{alpha}), we correct sign in the $\xi_{e\mathrm{L}}/2$ term compared to the misleading one in Ref.~\cite{Semikoz:2016lqv}. The thermodynamics term in the denominators in Eq.~(\ref{parameters}), $\tilde{p}+ \tilde{\rho}=2\pi^2g_*/45$, is given by the equation of state $p=\rho/3$ where $\rho=(2\pi^2/30)g_*T^4$ is the matter density.%, $g_*=106.75$ is the number of relativistic degrees of freedom in hot plasma.

Note that the effective magnetic diffusion coefficient $\tilde{\eta}_\mathrm{eff}$ and the helicity parameters $\tilde{\alpha}_\pm$ were studied in Ref.~\cite{Cam07}. The form of $\tilde{\alpha}_\pm$, obtained in Ref.~\cite{Cam07}, is different from that in Eq.~\eqref{parameters}. However, the numerical simulations (see Sec.~\ref{sec:BAUXIER} below) demonstrate that our main results are insensitive to this change of $\tilde{\alpha}_\pm$.

Let us stress that the vorticity $\bm{\omega}=\nabla\times {\bf v}$ does not contribute to the kinetic Eqs.~(\ref{spectra}) after substitution the drag velocity
in Eq.~(\ref{drag}) since the odd number of random HMFs appears in the kinetic equations for the spectra, $\mathcal{E}_{\rm B_\mathrm{Y}}$ and $\mathcal{H}_{\rm B_\mathrm{Y}}$. Namely, $\sim B_\mathrm{Y}^3$ comes from the term $\bm{\omega}$ for those binaries. Therefore, such a term vanishes after the application of the Wick theorem for the statistical averaging $\langle B_\mathrm{Y}B_\mathrm{Y}^*\dots\rangle$. The statistical averaging of multiple $B_\mathrm{Y}$-products is based on the canonical two-point correlator~\cite{Biskamp},
\begin{equation}\label{correlator}
  \langle
    B_i({\bf k},t)B_j({\bf p}, t)
  \rangle =
  \frac{(2\pi)^3}{2}\delta^{(3)}({\bf k} + {\bf p})
  \left[
    (\delta_{ij} - \hat{k}_i\hat{k}_j)S(k, t) + \mathrm{i} e_{ijk}\hat{k}_kA(k, t)
  \right].
\end{equation}
In Eq.~(\ref{correlator}), the form factors $S(k, t)$
and $A(k, t)$ are related to the spectra,
$\mathcal{E}_{\rm B_\mathrm{Y}}(k, t) = k^2S(k, t)/
(2\pi)^2$ and $\mathcal{H}_{\rm B_\mathrm{Y}}(k, t) = kA(k, t)/
(2\pi)^2$, obeying the kinetic Eq.~(\ref{spectra}) written there in conformal variables.

The limit $\tilde{k}_\mathrm{min}$ in the integrals in Eq.~(\ref{parameters}) depends on the inverse conformal time $\eta^{-1}$, or the inverse horizon size, being chosen as $\tilde{k}_\mathrm{min}=\eta^{-1}_0$ at the initial temparature $T_0$. The upper limit $\tilde{k}_\mathrm{max}$, $\tilde{k}_\mathrm{max}\gg \tilde{k}_\mathrm{min}$, is an arbitrary momentum at the lowest scale $L^{(\mathrm{min})}_{\mathrm{B}_\mathrm{Y}}=k_\mathrm{max}^{-1}$ for given HMFs. 

The kinetic Eq.~(\ref{spectra}) for the HMF spectra  should be solved self-consistently with the evolution equations for the fermion density asymmetries in the background matter, $\eta_f(t)=n_f(t) - n_{\bar{f}}(t)$.

\section{Kinetic equations for the particle density asymmetries in the symmetric phase\label{sec:KINEQS}} 
%of the early Universe, $T>T_\mathrm{EWPT}$}

The fermion density asymmetries in a hot uniform  plasma , $n_f(t) - n_{\bar{f}}(t)=T^3\xi_f(t)/6$, are characterized by the asymmetry parameter 
$\xi_f(t)$. We shall study its evolution in the conformal time, $\xi_f = \xi_f(\eta)$.

For the first generation of the lepton asymmetries $\xi_{e\mathrm{L}}(\eta)=\xi_{\nu_{e\mathrm{L}}}(\eta)$ and $\xi_{e\mathrm{R}}(\eta)$ accounting for the spin-flip $L\leftrightarrow R$ due to Higgs decays given by the rate~\cite{Davidson},
\begin{equation}\label{eq:spinfliprate}
  \Gamma(\eta)=2a\Gamma_\mathrm{RL}=
  \frac{242}{\eta_\mathrm{EW}}
  \left[
    1 -
    \left(
      \frac{\eta}{\eta_\mathrm{EW}}
    \right)^2
  \right],
  \quad
  \eta_\mathrm{EW}=\frac{\tilde{M}_\mathrm{Pl}}{T_\mathrm{EW}}=7\times 10^{15}.
\end{equation}
The sphaleron transition rate is $\Gamma_\mathrm{sph}=25\alpha_\mathrm{W}^5$, where $\alpha_\mathrm{W}=g^2/4\pi=\alpha_\mathrm{em}/\sin^2\theta_\mathrm{W}$ \cite{GR}. The system of the kinetic equations reads (compare with Eqs.~(13) and~(14) in Ref.~\cite{Semikoz:2016lqv}),
\begin{align}
  \label{right}
  \frac{{\rm d}\xi_{e\mathrm{R}}}{{\rm d}\eta}=&
  -\frac{3\alpha'}{\pi}\int_{\tilde{k}_\mathrm{min}}^{\tilde{k}_\mathrm{max}} \mathrm{d}\tilde{k}
  \frac{\partial \mathcal{\tilde{H}}_{\rm B_\mathrm{Y}}(\tilde{k},\eta)}{\partial\eta} -
  \Gamma(\eta)[\xi_{e\mathrm{R}}(\eta) -\xi_{e\mathrm{L}}(\eta) + \xi_0(\eta)],
  \\
  \label{left}
  \frac{{\rm d}\xi_{e\mathrm{L}}}{{\rm d}\eta}=&
  \frac{3\alpha'}{4\pi}\int_{\tilde{k}_\mathrm{min}}^{\tilde{k}_\mathrm{max}} \mathrm{d}\tilde{k}
  \frac{\partial \mathcal{\tilde{H}}_{\rm B_\mathrm{Y}}(\tilde{k},\eta)}{\partial \eta} -
  \frac{\Gamma(\eta)}{2}[\xi_{e\mathrm{L}}(\eta) -\xi_{e\mathrm{R}}(\eta) - \xi_0(\eta)] -
  \frac{\Gamma_\mathrm{sph}}{2}\xi_{e\mathrm{L}}(\eta),
  \\
  \label{higgs}
  \frac{{\rm d}\xi_0}{{\rm d}\eta}= &
  -\frac{\Gamma (\eta)}{2}
  \left[
    \xi_{e\mathrm{R}}(\eta)+ \xi_0(\eta) -\xi_{e\mathrm{L}}(\eta)
  \right].
\end{align}
Here we generalized the system of the kinetic equations derived in Ref.~\cite{Semikoz:2016lqv} by adding the kinetic Eq.~(\ref{higgs}) for the Higgs boson density asymmetry, $n_{\varphi^{(0)}} - n_{\tilde{\varphi}^{(0)}}=\xi_0T^3/3$, described by the parameter $\xi_0(\eta)=\mu_0/T$  when accounting for decays and inverse decays of the Higgs doublet, $H=(\varphi^{(+)},\varphi^{(0)})^{\rm T}$: $\bar{e}_\mathrm{R} e_\mathrm{L}\leftrightarrow \varphi^{(0)}$ and $\bar{e}_\mathrm{R}\nu_{e\mathrm{L}}\leftrightarrow \varphi^{(+)}$, as well as $e_\mathrm{R}\bar{e}_\mathrm{L}\leftrightarrow \tilde{\varphi}^{(0)}$ and $e_\mathrm{R}\nu_{e\mathrm{L}}\leftrightarrow \varphi^{(-)}$ for antiboson decays.

The `t~Hooft's law $B/3 - L=\text{const}$, where $L=L_{e\mathrm{R}} + L_{e\mathrm{L}} + L_{\nu_{e\mathrm{L}}}$ is the total lepton number in our scenario, provides the evolution of the baryon number $\mathrm{d}B/\mathrm{d}\eta$ given by the leptogenesis in the kinetic Eqs.~(\ref{right}) and~(\ref{left}). The corresponding solution for the baryon asymmetry $\text{BAU}(\eta)=(n_\mathrm{B} - n_{\bar{\mathrm{B}}})/s$ takes the form (see Eq.~(18) in Ref.~\cite{Semikoz:2016lqv}),
\begin{align}\label{BAU}
  \text{BAU}(\eta)=&
  5.3\times 10^{-3}\int_{\eta_0}^{\eta}\mathrm{d}\eta'
  \left\{
    \frac{{\rm d}\xi_{e\mathrm{R}}(\eta')}{{\rm d}\eta'} +
    \Gamma(\eta')[\xi_{e\mathrm{R}}(\eta') - \xi_{e\mathrm{L}}(\eta')]
  \right\}
  \nonumber
  \\
  & -
  \frac{6\times 10^7}{\eta_\mathrm{EW}}\int_{\eta_0}^{\eta}\xi_{e\mathrm{L}}(\eta')\mathrm{d}\eta'.
\end{align}
The matter noise influence on the BAU evolution in Eq.~(\ref{BAU}) is expected from the binary product for the HMF helicity density $\mathcal{\tilde{H}_{\rm B_\mathrm{Y}}}\sim {\bf Y}\cdot{\bf B}_\mathrm{Y}$ entering Eqs.~(\ref{right}) and~(\ref{left}).

\subsection{Generation of the chiral imbalance in the symmetric phase\label{sec:MU5}}

Subtracting Eq.~(\ref{left}) from Eq.~(\ref{right}), one gets the kinetic equation for the chiral imbalance $\tilde{\mu}_5(\eta)=\mu_5/T=[\xi_{e\mathrm{R}}(\eta) - \xi_{e\mathrm{L}}(\eta)]/2$,
 \begin{equation} \label{mu5}
  \frac{{\rm d}(\xi_{e\mathrm{R}}-\xi_{e\mathrm{L}})}{{\rm d}\eta}=
  - \frac{15\alpha'}{4\pi}\int_{\tilde{k}_\mathrm{min}}^{\tilde{k}_\mathrm{max}} \mathrm{d}\tilde{k}
  \frac{\partial\mathcal{\tilde{H}}_{\rm B_\mathrm{Y}}(\tilde{k},\eta)}{\partial\eta} -
  \frac{3\Gamma (\eta)}{2}\left(\xi_{e\mathrm{R}}-\xi_{e\mathrm{L}} + \xi_0\right) +
  \frac{\Gamma_\mathrm{sph}}{2}\xi_{e\mathrm{L}}.
\end{equation}
One can see that the important CME parameter $\tilde{\mu}_5$ depends explicitly on the Higgs boson asymmetry parameter  $\xi_0$ while the latter influences the baryon asymmetry in Eq.~(\ref{BAU}) through the lepton asymmetries only.

Let us remind that, in the absence of HMFs and corresponding Abelian abomalies for leptons, the equilibrium condition $\xi_{e\mathrm{R}} - \xi_{e\mathrm{L}} + \xi_0=0$ is implemented at $T\sim 1\,\mathrm{TeV}< T_\mathrm{RL}=T_0=10\,\mathrm{TeV}$ (see Eq.~(4.2) and Fig.~2(b) in Ref.~\cite{DS2}). In the presence of HMFs, this equilibrium fails, $\xi_{e\mathrm{R}} - \xi_{e\mathrm{L}} + \xi_0\neq 0$, and the boson asymmetry evolves as well, starting, e.g., from zero, $\xi_0^{(0)}=0$, and getting a negative value $\xi_0< 0$, as it should be for bosons.

\section{Complete set of the kinetic equations\label{sec:COMPLKINEQS}}

Collecting evolution equations for HMF spectra in Eq.~(\ref{spectra}), for the lepton and Higgs boson asymmetries in Eqs.~(\ref{right})-(\ref{higgs}), one obtains the complete system of self-consistent kinetic equations,
\begin{align}
  \frac{\partial\tilde{\mathcal{E}}_{\rm B_\mathrm{Y}}}{\partial\eta} = &
  -2\tilde{k}^{2}\eta_\mathrm{eff}\tilde{\mathcal{E}}_{\rm B_\mathrm{Y}}+
  \alpha_{+}\tilde{k}^{2}\tilde{\mathcal{H}}_{\rm B_\mathrm{Y}},
  \nonumber
  \\
  \frac{\partial\tilde{\mathcal{H}}_{\rm B_\mathrm{Y}}}{\partial\eta} = &
  -2\tilde{k}^{2}\eta_\mathrm{eff}\tilde{\mathcal{H}}_{\rm B_\mathrm{Y}}+
  4\alpha_{-}\tilde{\mathcal{E}}_{\rm B_\mathrm{Y}},
  \nonumber
  \\
  \frac{\mathrm{d}\xi_\mathrm{R}}{\mathrm{d}\eta} = &
  -\frac{3\alpha'}{\pi}\int_{\tilde{k}_\mathrm{min}}^{\tilde{k}_\mathrm{max}}\mathrm{d}\tilde{k}
  \frac{\partial\tilde{\mathcal{H}}_{\rm B_\mathrm{Y}}}{\partial\eta}-\Gamma(\xi_\mathrm{R}-\xi_\mathrm{L}+\xi_{0}),
  \nonumber
  \\
  \frac{\mathrm{d}\xi_\mathrm{L}}{\mathrm{d}\eta} = &
  \frac{3\alpha'}{4\pi}\int_{k_\mathrm{min}}^{\tilde{k}_\mathrm{max}}\mathrm{d}\tilde{k}
  \frac{\partial\tilde{\mathcal{H}}_{\rm B_\mathrm{Y}}}{\partial\eta}-
  \Gamma(\xi_\mathrm{L}-\xi_\mathrm{R}-\xi_{0})/2-
  \frac{\Gamma_\mathrm{sph}}{2}\xi_\mathrm{L},
  \nonumber
  \\
  \frac{\mathrm{d}\xi_{0}}{\mathrm{d}\eta} = &
  -\Gamma(\xi_\mathrm{R}+\xi_{0}-\xi_\mathrm{L})/2.
\end{align}
Here $\Gamma(T)=\Gamma_{0}(1-T_\mathrm{EW}^{2}/T^{2})$ is the rate of the chirality flip due to Higgs boson decays and $\Gamma_0=242/\eta_\mathrm{EW}$. In the radiation era, $t=(2{\rm H})^{-1}=\tilde{M}_\mathrm{Pl}/2T^2$, the conformal time $\eta$ can include any constant due to its definition $\mathrm{d}\eta=\mathrm{d}t/a=\mathrm{d}(\tilde{M}_\mathrm{Pl}/T)$ for the scale $a=1/T$.  We choose $\eta=M_{0}/T+\eta_{0}$ where $\eta_0=-\tilde{M}_\mathrm{Pl}/T_\mathrm{RL}=\text{const}$. Therefore, at the initial temperature $T_\mathrm{RL}=10\,\mathrm{TeV}$, we substitute the initial conformal time $\eta(T_\mathrm{RL})=0$, while at EWPT we substitute $\eta_\mathrm{EW}\approx \tilde{M}_\mathrm{Pl}/T_\mathrm{EW}=7\times 10^{15}$ since $T_\mathrm{EW} \ll T_\mathrm{RL}$. 

Then we introduce the new variables,
\begin{align}\label{eq:newvar}
  \tilde{\mathcal{E}}_{\rm B_\mathrm{Y}}(\tilde{k},\eta) = &
  \frac{\tilde{k}_\mathrm{max}\pi^{2}}{6\alpha'^{2}}R(\kappa,\tau),
  \quad
  \tilde{\mathcal{H}}_{\rm B_\mathrm{Y}}(\tilde{k},\eta)=
  \frac{\pi^{2}}{3\alpha'^{2}}H(\kappa,\tau),
  \quad
  \xi_\mathrm{R,L,0}(\tilde{\eta})=\frac{\pi\tilde{k}_\mathrm{max}}{\alpha'}M_\mathrm{R,L,0}(\tau),
  \nonumber
  \\
  \tau = &
  \frac{2\tilde{k}_\mathrm{max}^{2}}{\sigma_{c}}\eta,\quad\tilde{k}=\tilde{k}_\mathrm{max}\kappa,
\end{align}
where $\kappa_{m}<\kappa<1$, $\kappa_{m}=\tilde{k}_\mathrm{min}/\tilde{k}_\mathrm{max}$, and  $\tau\geq0$ is the new dimensionless time.

In new variables, we recast finally the complete system of kinetic equations in our problem:
\begin{align}\label{system3}
  \frac{\partial R}{\partial\tau} =&
  -\kappa^{2}
  \left(
    1+\frac{2\sigma_{c}\tilde{k}_\mathrm{max}^{2}\pi^{2}}{9\alpha'^{4}(\tilde{p}+\tilde{\rho})}
    \int_{\kappa_{m}}^{1}\mathrm{d}\kappa'R(\kappa',\tau)
  \right)R
  \nonumber
  \\
  & +
  \kappa^{2}
  \left(
    M_\mathrm{R}-\frac{M_\mathrm{L}}{2}-
    \frac{2\sigma_{c}\tilde{k}_\mathrm{max}^{2}\pi^{2}}{9\alpha'^{4}(\tilde{p}+\tilde{\rho})}
    \int_{\kappa_{m}}^{1}\mathrm{d}\kappa'\kappa'^{2}H(\kappa',\tau)
  \right)H,
  \nonumber
  \\
  \frac{\partial H}{\partial\tau} =&
  -\kappa^{2}
  \left(
    1+\frac{2\sigma_{c}\tilde{k}_\mathrm{max}^{2}\pi^{2}}{9\alpha'^{4}(\tilde{p}+\tilde{\rho})}
    \int_{\kappa_{m}}^{1}\mathrm{d}\kappa'R(\kappa',\tau)
  \right)H
  \nonumber
  \\
  & +
  \left(
    M_\mathrm{R}-\frac{M_\mathrm{L}}{2}+
    \frac{2\sigma_{c}\tilde{k}_\mathrm{max}^{2}\pi^{2}}{9\alpha'^{4}(\tilde{p}+\tilde{\rho})}
    \int_{\kappa_{m}}^{1}\mathrm{d}\kappa'\kappa'^{2}H(\kappa',\tau)
  \right)R,
  \nonumber
  \\
  \frac{\mathrm{d}M_\mathrm{R}}{\mathrm{d}\tau} =&
  \int_{\kappa_{m}}^{1}\mathrm{d}\kappa'\kappa'^{2}H(\kappa',\tau)-
  \left(
    M_\mathrm{R}-\frac{M_\mathrm{L}}{2}
  \right)
  \int_{\kappa_{m}}^{1}\mathrm{d}\kappa'R(\kappa',\tau)
  \notag
  \\
  & -
  \Gamma'(M_\mathrm{R}-M_\mathrm{L}+M_{0}),
  \nonumber
  \\
  \frac{\mathrm{d}M_\mathrm{L}}{\mathrm{d}\tau} =&
  -\frac{1}{4}\int_{\kappa_{m}}^{1}\mathrm{d}\kappa'\kappa'^{2}H(\kappa',\tau)+
  \frac{1}{4}
  \left(
    M_\mathrm{R}-\frac{M_\mathrm{L}}{2}
  \right)
  \int_{\kappa_{m}}^{1}\mathrm{d}\kappa'R(\kappa',\tau)
  \nonumber
  \\
  & -
  \Gamma'(M_\mathrm{L}-M_\mathrm{R}-M_{0})/2-\frac{\Gamma'_{s}}{2}M_\mathrm{L},
  \nonumber
  \\
  \frac{\mathrm{d}M_{0}}{\mathrm{d}\tau} =&
  -\Gamma'(M_\mathrm{R}+M_{0}-M_\mathrm{L})/2,
\end{align}
where $\Gamma'_{s}=\sigma_{c}\Gamma_\mathrm{sph}/2\tilde{k}_\mathrm{max}^{2}$,
\begin{equation}\label{chiral}
  \Gamma'(\tau)=\frac{\sigma_{c}\Gamma_{0}}{2\tilde{k}_\mathrm{max}^{2}}
  \left(
    1-\frac{T_\mathrm{EW}^{2}}{T_\mathrm{RL}^{2}}
    \left[
      1+\frac{T_\mathrm{RL}}{M_{0}}\frac{\sigma_{c}}{2\tilde{k}_\mathrm{max}^{2}}\tau
    \right]^{2}
  \right),
\end{equation}
and $\Gamma_{0} = 242/\eta_\mathrm{EW}$.

%Note that
%\begin{equation}\label{helicity1}
%H_{0}(\kappa)=qR_{0}(\kappa)/\kappa.
%\end{equation}
Then we recast BAU in Eq.~(\ref{BAU}) in these new variables,
\begin{align}\label{BAU2}
  \mathrm{BAU}(\tau)= &
  1.76\times\tilde{k}_\mathrm{max}\int_0^{\tau}\mathrm{d}\tau'
  \left\{
    \frac{{\rm d}M_\mathrm{R}}{{\rm d}\tau'} + \Gamma'(\tau')
    \left[
      M_\mathrm{R}(\tau')- M_\mathrm{L}(\tau')
    \right]
  \right\}
  \nonumber
  \\
  & -
  \frac{10^{-4}}{\tilde{k}_\mathrm{max}}\int_0^{\tau}M_\mathrm{L}(\tau')\mathrm{d}\tau',
\end{align}
where the rate of the chirality flip due to Higgs decays, $\Gamma'(\tau')$, is given by Eq.~(\ref{chiral}). One can see in Eq.~(\ref{BAU2}) that BAU increases for small-scale (random) HMFs, when $\tilde{k}_\mathrm{max}$ increases. The sphaleron transition influence in last line of Eq.~(\ref{BAU2}) reduces with a growth of $\tilde{k}_\mathrm{max}$.

Finally we rewrite the kinetic Eq.~(\ref{mu5}) for the chiral imbalance $M_\mathrm{R} - M_\mathrm{L}\sim \mu_5$,
\begin{equation}\label{mu5new}
  \frac{{\rm d}}{{\rm d}\tau}(M_\mathrm{R} - M_\mathrm{L})=
  -\frac{5}{4}\int_{\kappa_m}^1\mathrm{d}\kappa\frac{\partial H(\tau,\kappa)}{\partial \tau} -
  \frac{3}{2}\Gamma'(M_\mathrm{R} - M_\mathrm{L} + M_0) + \Gamma_s'\frac{M_\mathrm{L}}{2},
\end{equation}
that should be solved self-consistently with the all kinetic equations in (\ref{system3}) including that for the Higgs boson asymmetry $M_0$.

\subsection{Initial condition\label{sec:INICOND}}

We choose the same initial conditions from Eqs.~(9)-(11) in Ref.~\cite{Semikoz:2016lqv}. Namely,
\begin{equation}\label{helicity0}
  \tilde{\mathcal{E}}(\tilde{k},0) \sim \tilde{k}^{n_{\rm B_\mathrm{Y}}},
  \quad
  \tilde{\mathcal{H}}(\tilde{k},0) = 2q \tilde{\mathcal{E}}(\tilde{k},0)/\tilde{k},
  \quad
  0\leq q\leq 1.
\end{equation}
The initial condition in Eq.~\eqref{helicity0} can be rewritten in the new variables in Eq.~\eqref{eq:newvar} as
\begin{equation}\label{helicity}
  H(\kappa,0)=q\frac{R(\kappa,0)}{\kappa}.
\end{equation}
Adopting the initial HMF Kolmogorov energy density spectrum, one gets that
\begin{equation}\label{energy}
  R(\kappa, 0)=A\kappa^{n_{\rm B_\mathrm{Y}}}, 
  \quad
  n_{\rm B_\mathrm{Y}}= -5/3,
\end{equation}
where the normalization constant $A$,
\begin{equation}
  A=\frac{3\alpha'^2(1 + n_{\rm B_\mathrm{Y}})[\tilde{B}^{(0)}_\mathrm{Y}]^2}
  {\pi^2\tilde{k}_\mathrm{max}^2(1 - \kappa_m^{1 + n_{\rm B_\mathrm{Y}}})},
\end{equation}
and
\begin{equation}
  \tilde{B}^{(0)}_\mathrm{Y}=
  \left[
    2\int_{\tilde{k}_\mathrm{min}}^{\tilde{k}_\mathrm{max}} \mathrm{d}\tilde{k}\mathcal{\tilde{E}}_{{\rm B_\mathrm{Y}}}(\tilde{k},0)
  \right]^{1/2}
\end{equation}
is a seed HMF. The parameter
$\tilde{k}_\mathrm{min}=\tilde{l}_\mathrm{H}^{-1}(\tau=0)\simeq 10^{-14}$ corresponds to the maximum scale of HMF at $\eta=\tau=0$, whereas $\tilde{k}_\mathrm{max}$ is an arbitrary wave number $\tilde{k}_\mathrm{max}\gg \tilde{k}_\mathrm{min}$ in our causal scenario. 

Other parameters, entering the kinetic equations, have the following values: $\alpha'=g'^{2}/4\pi=\alpha_\mathrm{em}/\cos^2\theta_\mathrm{W}=9.5\times 10^{-3}$, $\Gamma_\mathrm{sph}=8\times 10^{-7}$ is the dimensionless rate of sphaleron transitions, i.e. the rate multiplied by the temperature $T$. The dimensionless spin-flip rate $\Gamma(\eta)$ in Eq.~\eqref{eq:spinfliprate} is very small, $\Gamma(\eta)\ll \Gamma_\mathrm{sph}$, because of the multiplier, $242/\eta_\mathrm{EW}=3.4\times 10^{-14}$. However, in the kinetic equations, the huge sphaleron transition rate is multiplied by a small $\xi_{e\mathrm{L}}(\eta)$, for which we choose the following initial conditions,
\begin{equation}\label{start}
  \xi_{e\mathrm{L}}(\tau=0)=\xi_0(\tau=0)={\rm BAU}(\tau=0)=0.
\end{equation}
We can start from the non-zero right electron asymmetry $\xi_{e\mathrm{R}}(\tau=0)=10^{-10}$, or $M_\mathrm{R}(\tau=0)=(\alpha'/\pi\tilde{k}_\mathrm{max})10^{-10}$ at the level of BAU expected at $\eta_\mathrm{EW}=7\times 10^{15}$. In Ref.~\cite{Semikoz:2016lqv}, we probed $\xi_{e\mathrm{R}}(\eta_0)$ in the range from a very small $\xi_{e\mathrm{R}}(\eta_0)=10^{-14}$ up to $\xi_{e\mathrm{R}}(\eta_0)=10^{-6}$.
 
\section{Matter noise influences BAU and $\xi_{e\mathrm{R}}$ evolutions\label{sec:BAUXIER}}

We claim in Sec.~\ref{sec:INTR}, that BAU results from nonzero HMFs via the Abelian anomaly for $e_\mathrm{R}$ as pointed in Refs.~\cite{Shaposh1,Shaposh2}. The matter turbulence can influence BAU in a different way. These matter perturbations are connected in the symmetric phase with stong HMFs.

The strong seed HMFs $B_\mathrm{Y}^{(0)}=1.4 \times 10^{-2} T_0^2$ and $B_\mathrm{Y}^{(0)}=1.4 \times 10^{-1} T_0^2$ are used in our numerical simulations.\footnote{Here the initial temperature $T_0=10\,\mathrm{TeV}$ corresponds to the HMF strength $T^2_0=5\times 10^{27}\,\mathrm{G}$} Such HMFs  do not violate Friedmann law, since their HMF energy densities obey $B^2_\mathrm{Y}/2\ll \rho=g_*\pi^2T^4/30$ . For sure these HMFs do not violate the BBN limit on primordial magnetic fields, derived in Ref.~\cite{Schramm}, $B_\mathrm{BBN}\leq 10^{11}\,\mathrm{G}$, at the temperature $T_\mathrm{BBN}\sim 0.1\,\mathrm{MeV}$, since $T_0/T_\mathrm{BBN}\simeq 10^8$ or $B_\mathrm{Y}\sim 10^{27}\,\mathrm{G}$ for $[T_0/T_\mathrm{BBN}]^2\simeq 10^{16}$. 

\begin{figure}
  \centering
  \subfigure[]
  {\label{fig:BAUa}
  \includegraphics[scale=.39]{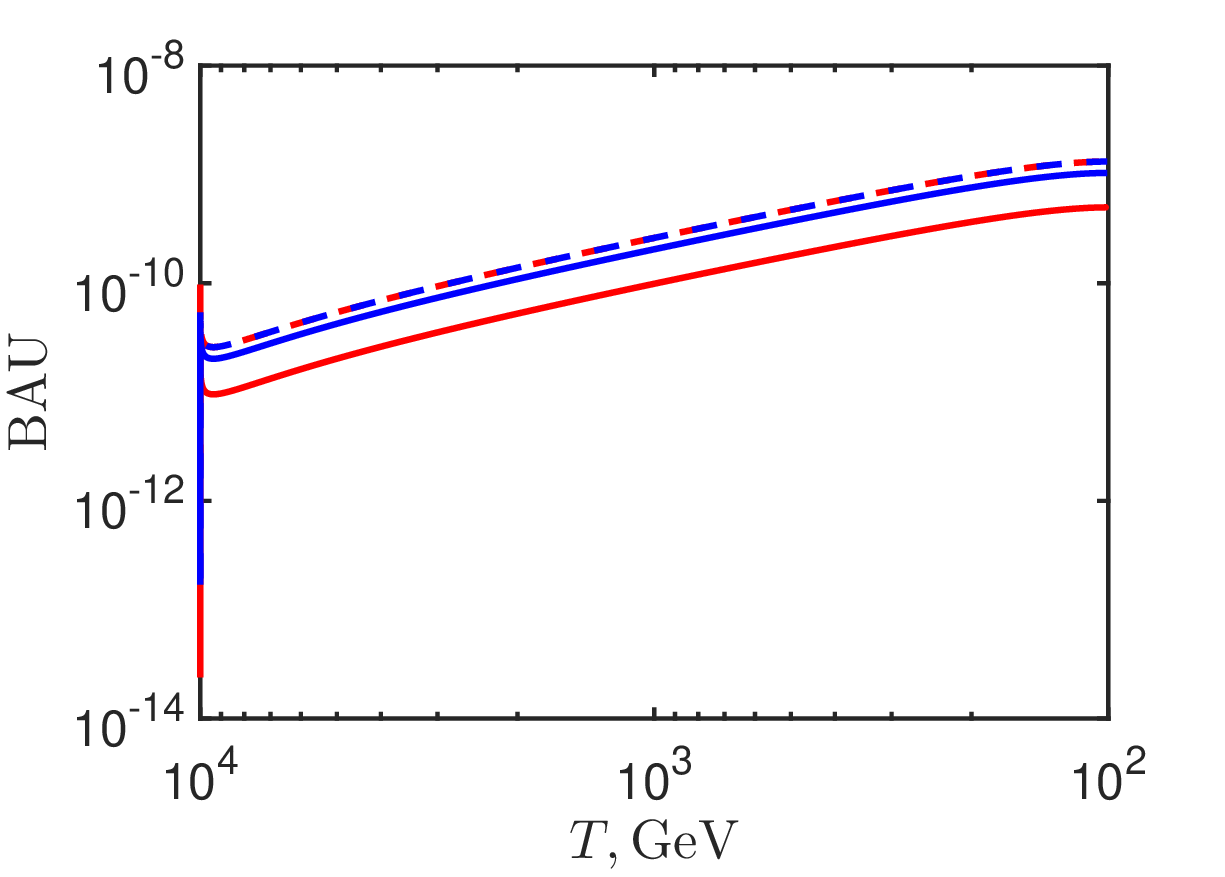}}
  \hskip-.6cm
  \subfigure[]
  {\label{fig:BAUb}
  \includegraphics[scale=.39]{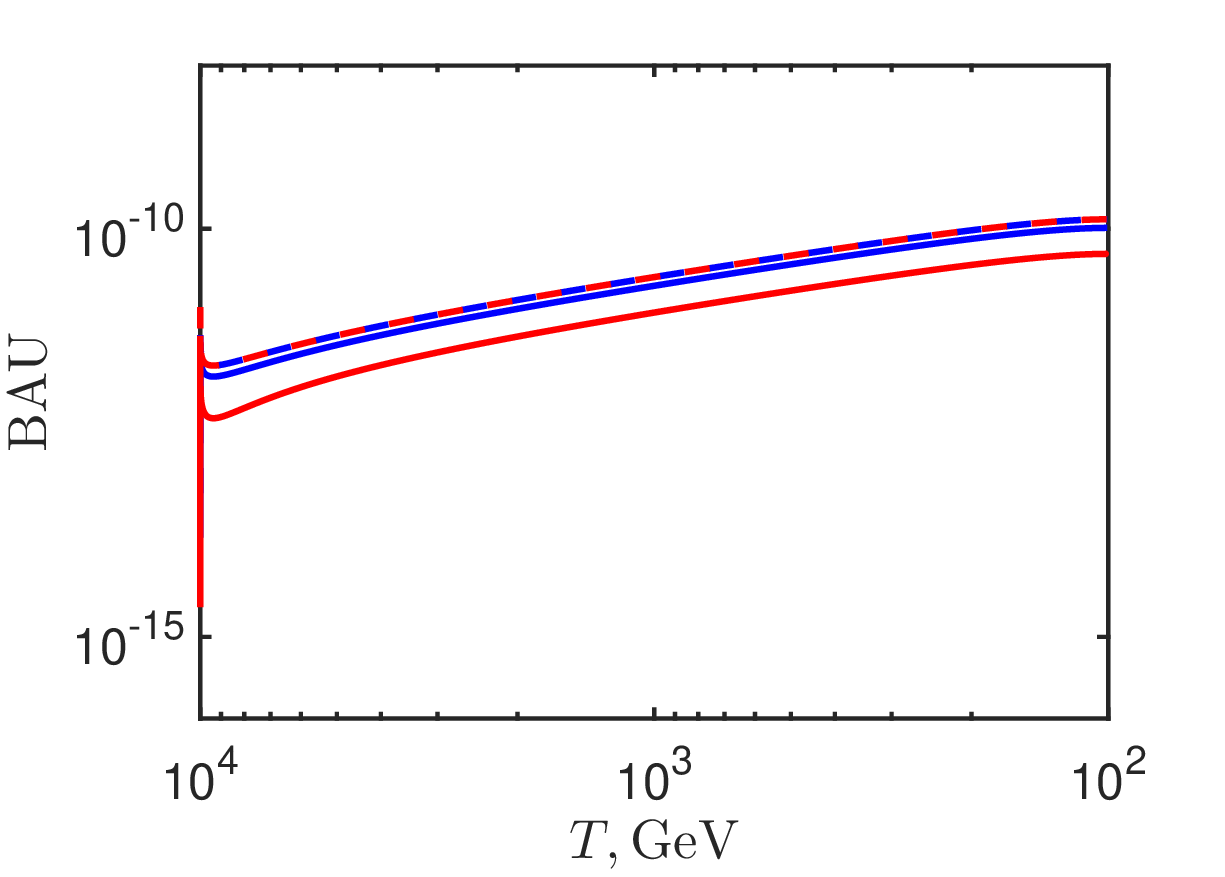}}
  \\
  \subfigure[]
  {\label{fig:BAUc}
  \includegraphics[scale=.39]{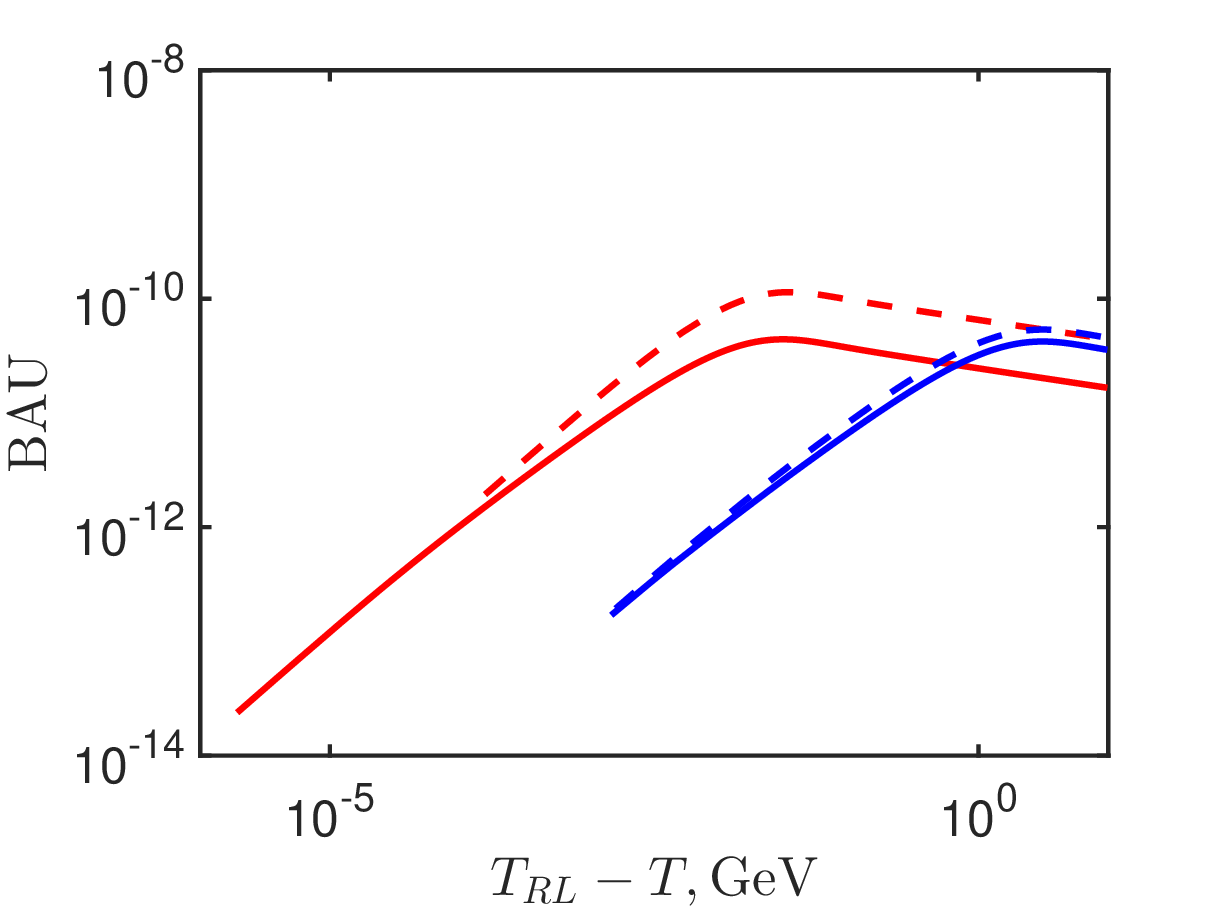}}
  \hskip-.6cm
  \subfigure[]
  {\label{fig:BAUd}
  \includegraphics[scale=.39]{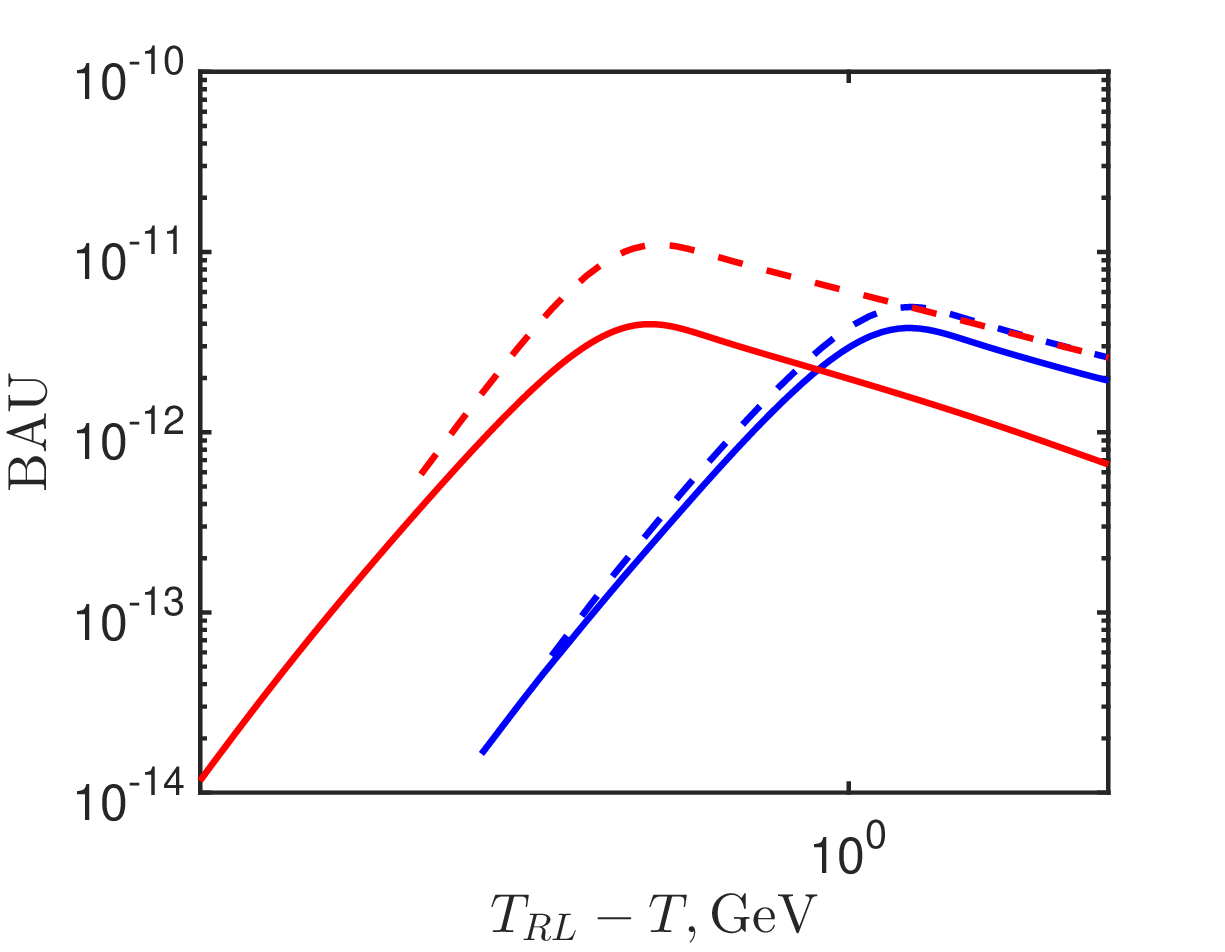}}
  \protect
  \caption{BAU growth in the symmetric phase within the temperature range
  $T_\mathrm{RL}=10\,{\rm TeV}\leq T\leq T_\mathrm{EW}=100\,\text{GeV}$ for
  $\tilde{k}_\mathrm{max}=10^{-3}$, $\xi_{e\mathrm{R}}^{(0)}=10^{-10}$, $\xi_{0}^{(0)}=0$, and $\xi_{e\mathrm{L}}^{(0)}=0$.
  Solid lines account for the noise contribution, whereas dashed lines are without noise.
  Red lines correspond to the initial HMF strength $\tilde{B}_\mathrm{Y}^{(0)}=1.4 \times 10^{-1}$ or $B_\mathrm{Y}^{(0)}=7\times 10^{26}~\,{\rm G}$
  and blue lines correspond to $\tilde{B}_\mathrm{Y}^{(0)}=1.4 \times 10^{-2}$ or $B_\mathrm{Y}^{(0)}=7\times 10^{25}\,{\rm G}$. The curves
  (a) correspond to the fully helical HMF in Eq.~(\ref{helicity}), $q=1$; in panel (b) the less helicity $q=0.1$ was substituted.
  (c) The same as in panel (a) for small evolution times $10^{-6}\,\text{GeV} < T_\mathrm{RL} - T < 10\,\text{GeV}$.
  (d) The same as in panel (b) for small evolution times $10^{-5}\,\text{GeV} < T_\mathrm{RL} - T < 10^2\,\text{GeV}$.
  \label{fig:BAU}}
\end{figure}

\begin{figure}
  \centering
  \subfigure[]
  {\label{fig:xieRa}
  \includegraphics[scale=.39]{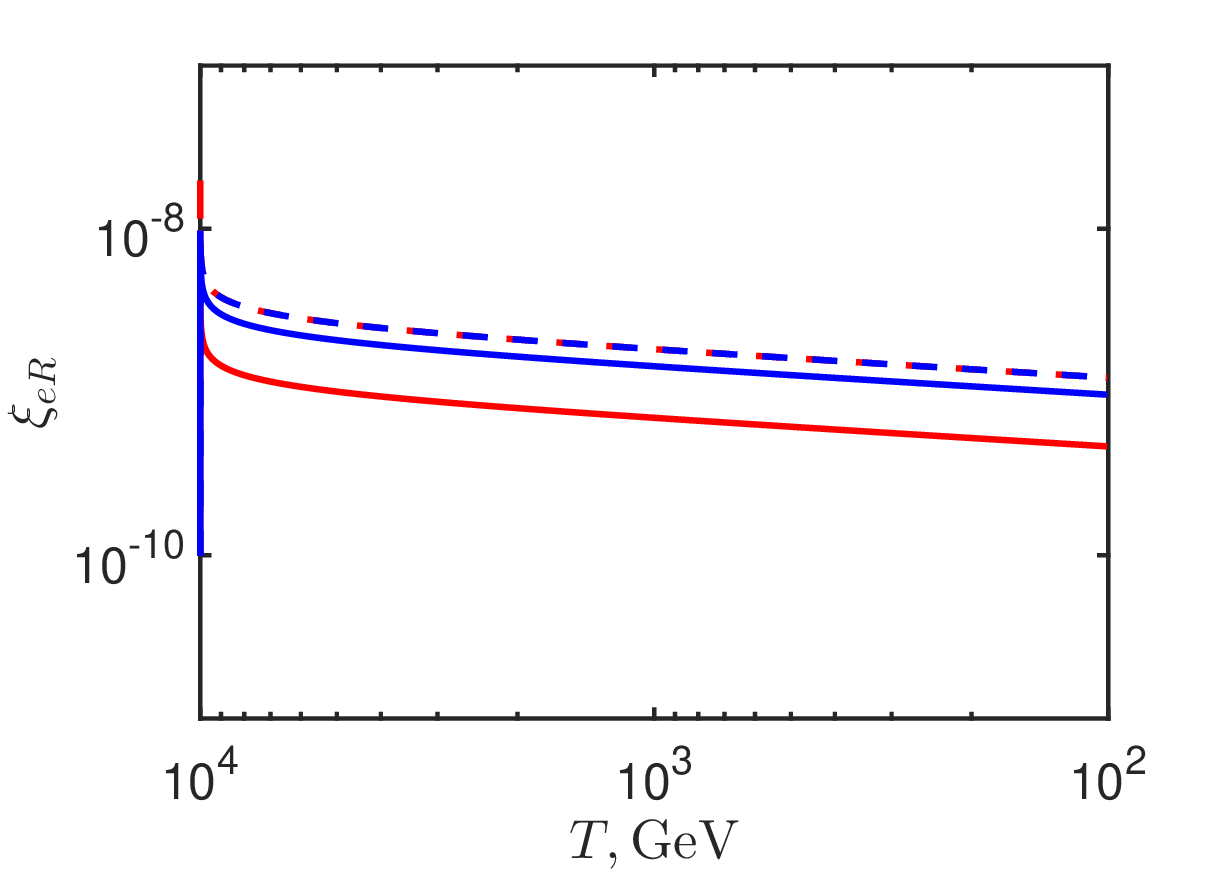}}
  \hskip-.6cm
  \subfigure[]
  {\label{fig:xieRb}
  \includegraphics[scale=.39]{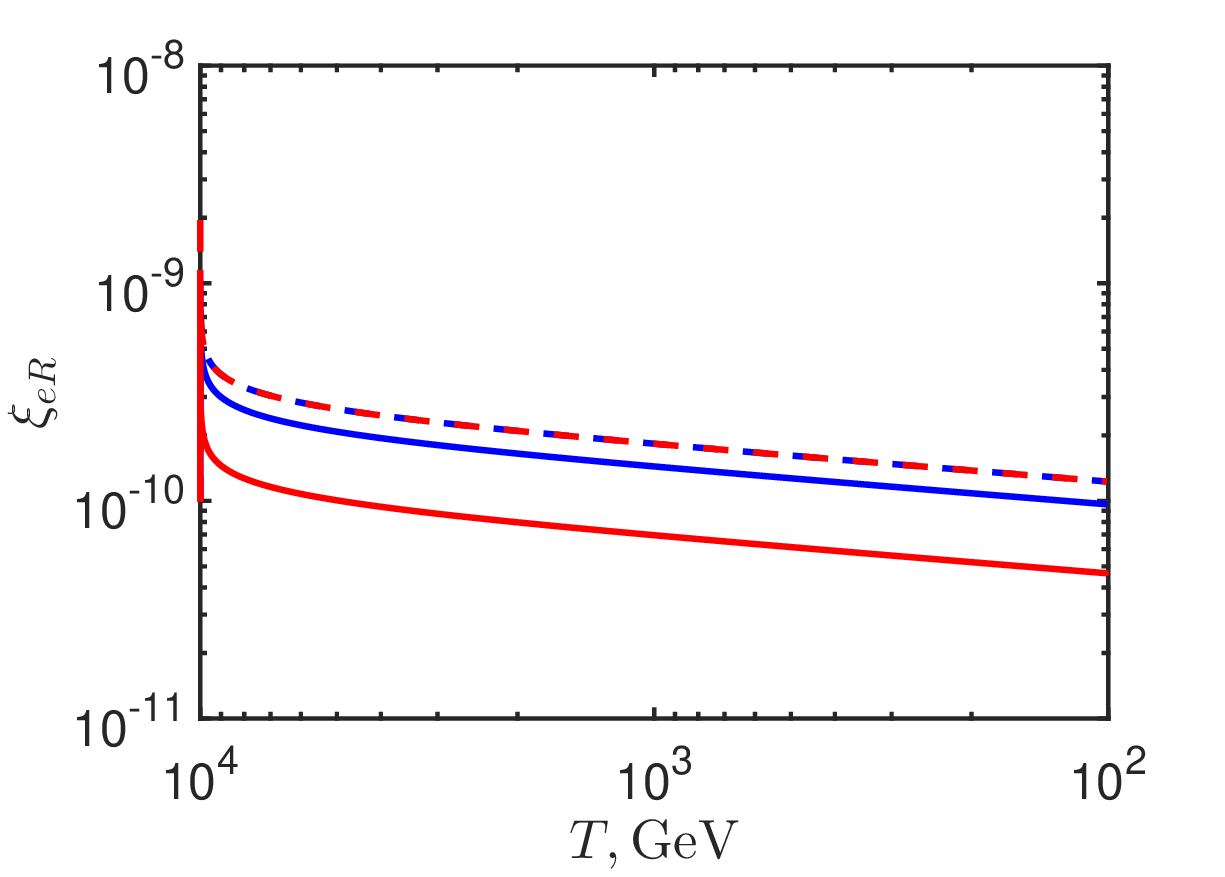}}
  \\
  \subfigure[]
  {\label{fig:xieRc}
  \includegraphics[scale=.39]{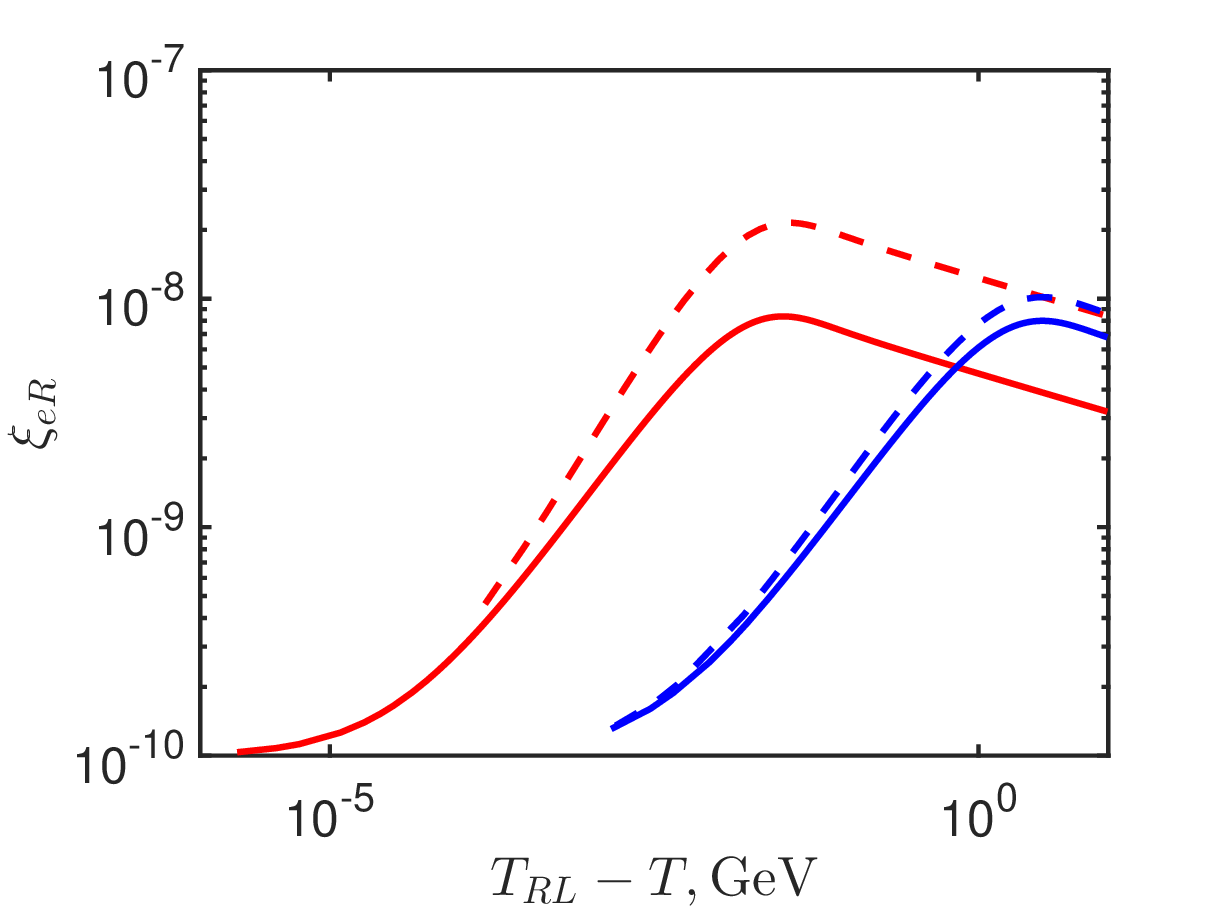}}
  \hskip-.6cm
  \subfigure[]
  {\label{fig:xieRd}
  \includegraphics[scale=.39]{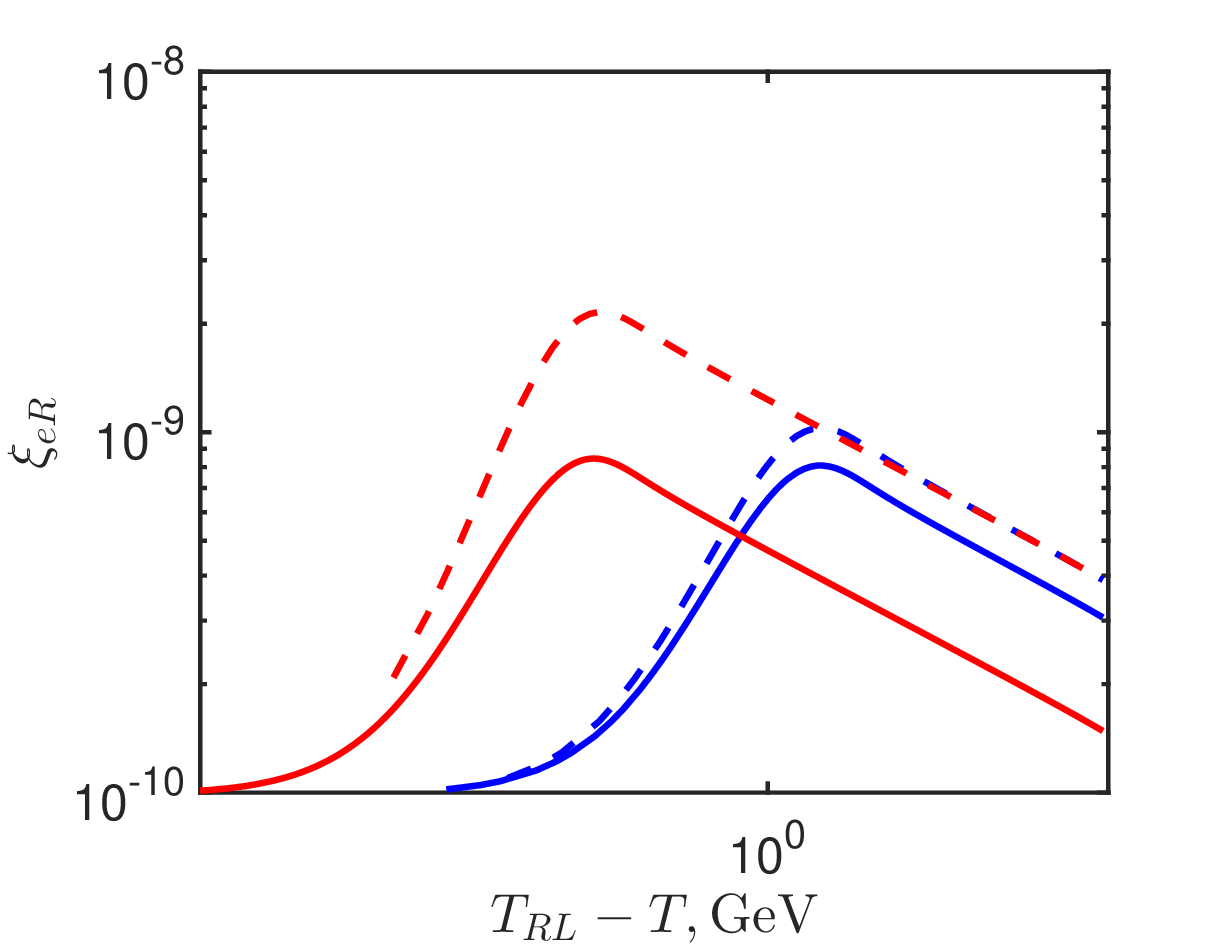}}
  \protect
  \caption{The evolution of the right electrons asymmetry in the temperature range
  $T_\mathrm{RL}=10\,{\rm TeV}\leq T\leq T_\mathrm{EW}=100\,\text{GeV}$ for
  $\tilde{k}_\mathrm{max}=10^{-3}$, $\xi_{e\mathrm{R}}^{(0)}=10^{-10}$, $\xi_{0}^{(0)}=0$, and $\xi_{e\mathrm{L}}^{(0)}=0$.
  Solid lines account for the noise contribution, wheres dashed lines are without noise.
  Red lines correspond to $\tilde{B}_\mathrm{Y}^{(0)}=1.4 \times 10^{-1}$ or $B_\mathrm{Y}^{(0)}=7\times 10^{26}~\,{\rm G}$
  and blue lines correspond to $\tilde{B}_\mathrm{Y}^{(0)}=1.4 \times 10^{-2}$ or $B_\mathrm{Y}^{(0)}=7\times 10^{25}\,{\rm G}$.
  The curves (a) correspond to the fully helical HMF in Eq.~(\ref{helicity}), $q=1$; in panel (b) the less helicity $q=0.1$ was substituted.
  (c) The same as in panel (a) for small evolution times $10^{-6}\,\text{GeV} < T_\mathrm{RL} - T < 10\,\text{GeV}$.
  (d) The same as in panel (b) for small evolution times $10^{-5}\,\text{GeV} < T_\mathrm{RL} - T < 10^2\,\text{GeV}$.
  \label{fig:xieR}}
\end{figure}

Firstly, a matter turbulence caused by random HMFs  diminishes all fermion asymmetries including the BAU in Fig.~\ref{fig:BAU} and the right lepton asymmetry $\xi_{e\mathrm{R}}$ in Fig.~\ref{fig:xieR}. The decline of the fermion asymmetries when accounting for a matter noise are seen in all plots below. 

In Figs.~\ref{fig:BAUa} and~\ref{fig:BAUc}, the fully helical HMFs with $q=1$ provide a greater increase of BAU comparing with a small $q=0.1$ in Figs.~\ref{fig:BAUb} and~\ref{fig:BAUd}, for which BAU reduces at least one order of magnitude. This BAU increase is especially noticeable near $T_\mathrm{RL}=10^4\,\mathrm{GeV}$; cf. Figs.~\ref{fig:BAUc} and~\ref{fig:BAUd}. It is also remarkable that noiseless (dashed) curves in Figs.~\ref{fig:BAUc} and~\ref{fig:BAUd} begin coincide for early times $\sim T_\mathrm{RL}-T\ll T_\mathrm{RL}$ near the ends of curves , during a few tens ${\rm GeV}$ for the initial HMFs $B_\mathrm{Y}^{(0)}=7\times 10^{25}\,\mathrm{G}$ and $B_\mathrm{Y}^{(0)}=7\times 10^{26}\,\mathrm{G}$.

The similar dependence in the evolution of noiseless (dashed) curves and solid (blue and red) curves under a matter turbulence is seen in Fig.~\ref{fig:xieR} for the  right electron asymmetry $\xi_{e\mathrm{R}}$. Of course, for the higher helicity $q=1$ in Figs.~\ref{fig:xieRa} and \ref{fig:xieRc}, the asymmetry $\xi_{e\mathrm{R}}$ is one order of magnitude bigger comparing that  for $q=0.1$ in Figs.~\ref{fig:xieRb} and~\ref{fig:xieRd}.

We have also explored the evolution of $\xi_{e\mathrm{L}}$ versus $T$ for the same initial condition as in Fig.~\ref{fig:xieR}. It turns out to be much less than $\xi_{e\mathrm{R}}$ for all range of the temperature variation, as it was predicted in Ref.~\cite{DS2}. Therefore, we omit the corresponding results.

\subsection{Scales of strong $B_\mathrm{Y}$ relevant to matter turbulence and sensitive to ohmic losses}

The noise matter influence becomes noticeable at sufficiently strong HMFs correspondingly to their small-scale values $L^{(\mathrm{min})}_{\mathrm{B}_\mathrm{Y}}=k_\mathrm{max}^{-1}$  relevant to the ohmic diffusion for such hypermagnetic fields. We remind that large-scale HMFs, $L_{\mathrm{B}_\mathrm{Y}}\gg  L^{(\mathrm{min})}_{\mathrm{B}_\mathrm{Y}}$, survive versus ohmic dissipation.
Then we consider non-dissipative HMFs for which even a minimal scale $L_{\mathrm{B}_\mathrm{Y}}^{(\mathrm{min})}$ is large enough, $L^{(\mathrm{min})}_{\mathrm{B}_\mathrm{Y}}\geq (2H\sigma_\mathrm{cond})^{-1/2}$. The latter criterion follows from the obvious comparison of the expansion time during the radiation era, $t=(2\mathrm{H})^{-1}$, and the ohmic diffusion time $t_\mathrm{diff}=\sigma_\mathrm{cond}[L^{(\mathrm{min})}_{\mathrm{B}_\mathrm{Y}}]^2$, $t<t_\mathrm{diff}$, or $(2H)^{-1}\leq \sigma_\mathrm{cond}[L^{(\mathrm{min})}_{\mathrm{B}_\mathrm{Y}}]^2$. Substituting $\sigma_\mathrm{cond}=100T$ we obtain the upper bound $k_\mathrm{max}\leq \sqrt{200T^3/\tilde{M}_\mathrm{Pl}}$, or $\tilde{k}_\mathrm{max}\leq \sqrt{200/\eta}$. The latter criterion for the survival of HMFs changes within the range $\tilde{k}_\mathrm{max}\leq (10^{-7} - 10^{-6})$ correspondingly to  conformal times $\eta_\mathrm{EW} > \eta_0$ in the symmetric phase. 

Thus, for a wide maximum range $\tilde{k}_\mathrm{max}=(10^{-2}- 10^{-5})$ for which  we selected the only momentum $\tilde{k}_\mathrm{max}=10^{-3}$ in our plots, the ohmic dissipation is presented for HMFs together with the matter noise influence. For a smaller $\tilde{k}_\mathrm{max}\leq (10^{-7} - 10^{-6})$, the noise matter disappears while HMFs survive versus ohmic diffusion.

\section{Discussion\label{sec:DISC}}

We consider a novel scenario for the BAU generation in the symmetric phase in the early Universe, for which random HMFs, $\langle {\bf B}_\mathrm{Y}\rangle =0$, being small-scale at distances $r_\mathrm{D}\simeq 10/T\ll L^{(\mathrm{min})}_{\mathrm{B}_\mathrm{Y}}\simeq 10^3/T$, where $r_\mathrm{D}$ is the Debye radius, dominate and motivate a turbulent matter involvement. Conversely to small-scale distances $\sim L^{(\mathrm{min})}_{\mathrm{B}_\mathrm{Y}}$ the majority of matter spread upon a rather large-scale HMFs, $L^{(\mathrm{min})}_{\mathrm{B}_\mathrm{Y}}\ll L^{(\mathrm{max})}_{\mathrm{B}_\mathrm{Y}}\leq 10^{14}/T$, within the temperature region $T\leq T_0$ does not feel a turbulent motion at all.

This situation resembles the representation of a magnetic field in a star in the standard magnetic hydrodynamics (MHD), ${\bf H}= {\bf B} + {\bf b}$, combined from the mean field term ${\bf B}$ and a random field ${\bf b}$, obeying the condition $\langle {\bf b}\rangle=0$. The random magnetic field ${\bf b}$ evidently gets a strong value in its amplitude, $\sqrt{b^2} >  B$, for example, within the convection zone in the Sun (see, e.g., Ref.~\cite{Sti04}). The random HMF ${\bf B}_\mathrm{Y}$ is also quite strong in the considered problem and can be treated here as the analogue for ${\bf b}$ in the standard MHD.

We considered the turbulent matter appearance through the dynamo term $\nabla\times ({\bf v}\times {\bf B}_\mathrm{Y})$ entering the induction Eq.~(\ref{Faraday}) due to the change of a such an advection contribution via the Lorentz force which dominates over the fluid velocity in Eq.~(\ref{drag}). Certainly, for the evolution of the binary spectra $\rho_{\mathrm{B}_\mathrm{Y}}\sim B_\mathrm{Y}^2$ and $h_\mathrm{Y}\sim {\bf Y}\cdot{\bf B}_\mathrm{Y}$, the additional terms that are proportional to $\sim B_\mathrm{Y}^4$ arise originally due to advection term $\nabla\times ({\bf v}\times {\bf B}_\mathrm{Y})\sim B_\mathrm{Y}^3$ in the Faraday Eq.~(\ref{Faraday}). Then, a such term, $\sim B_\mathrm{Y}^3$, is multiplied by  ${\bf B}_\mathrm{Y}^*$ leading to the evolution of the binary spectra.  Namely, these matter noise terms $\sim B_\mathrm{Y}^4$ appear in the evolution Eq.~(\ref{spectra}). They arise due to the additional  integrals $\sim \int_{\tilde{k}_\mathrm{min}}^{\tilde{k}_\mathrm{max}} \mathrm{d}\tilde{k}\mathcal{\tilde{E}}_{{\rm B_\mathrm{Y}}}$ and $\sim \int_{\tilde{k}_\mathrm{min}}^{\tilde{k}_\mathrm{max}} \mathrm{d}\tilde{k}\tilde{k}^2\mathcal{\tilde{H}}_{{\rm B_\mathrm{Y}}}$ presented in the parameters in Eq.~(\ref{parameters}), which compete there with noiseless conductivity terms $\sigma_c^{-1}$ in $\eta_\mathrm{eff}$ and $\alpha_\mathrm{Y}(\eta)\sim \sigma_c^{-1}$ in the coefficients $\alpha_{\pm}$. 

Finally, these matter turbulence terms $\sim B_\mathrm{Y}^4$ crucially depend on a range for a varying limit $k_\mathrm{max}=[L^{(\mathrm{min})}_{\mathrm{B}_\mathrm{Y}}]^{-1}$ discussed in the beginning of this section. In the numerical simulations, we selected $\tilde{k}_\mathrm{max}=10^{-3}$ to feel difference between the matter turbulence influencing the BAU generation and a noiseless regime neglecting such turbulence as shown by dashed lines in Figs.~\ref{fig:BAU} and~\ref{fig:xieR}.

To resume, for the first time, we have considered the matter turbulence before the EWPT, $T> T_\mathrm{EW}$, caused by strong HMFs in the symmetric phase in the early Universe. This matter noise diminishes fermion asymmetries including the BAU generation and may be crucial for a future refinement of the simplest SM model suggested here.

\section*{Acknowledgements}
We acknowledge D.~D.~Sokoloff and A.~Yu.~Smirnov for useful discussions.


\begin{thebibliography}{100}
\bibitem{fn1}
  The drag time is estimated as the free path
  $l_\mathrm{free}=(n\sigma)^{-1}\approx 1/\alpha'^{2}T$,
  where $n\sim T^3$ is the hot plasma density and
  $\sigma \sim \alpha'^{2}/T^2$ is the Coulomb cross section for
  all leptons including neutrinos which have the hypercharge as well.
\end{thebibliography}
\end{document}